\documentclass[aps,prb,twocolumn,superscriptaddress,floatfix]{revtex4}

\usepackage{graphicx}

\begin{document}

\title{Aging-induced complex transformation behavior of martensite
in Ni$_{57.5}$Mn$_{17.5}$Ga$_{25}$ shape memory alloy}

\author{V.V.~Khovailo}
\affiliation{National Institute of Advanced Industrial Science and
Technology, Tohoku Center, Sendai 983-8551, Japan}

\author{R.~Kainuma}
\affiliation{Department of Materials Science, Graduate School of
Engineering, Tohoku University, Sendai 980-8579, Japan}

\author{T.~Abe}
\author{K.~Oikawa}
\affiliation{National Institute of Advanced Industrial Science and
Technology, Tohoku Center, Sendai 983-8551, Japan}

\author{T.~Takagi}
\affiliation{Institute of Fluid Science, Tohoku University, Sendai
980-8577, Japan}

\begin{abstract}

Ni$_{57.5}$Mn$_{17.5}$Ga$_{25}$ shape memory alloy exhibits a
complex transformation behavior, appearing after aging. Aging in
the austenitic state resulted in an ordinary decrease of the
martensitic transformation temperature. Contrary to this, aging in
the martensitic state brought about unusual features of the
martensitic transformation observed so far only in Ni-Ti alloys.

\end{abstract}

\maketitle

\section{Introduction}

Early studies of ferromagnetic shape memory alloys Ni-Mn-Ga have
already indicated that the martensitic transformation temperature
is sensitive to composition and can be observed in a wide
temperature range~\cite{1-c}. Further systematic studies of
composition dependencies of the martensitic transformation
temperature $T_m$ showed that there is a relation between $T_m$
and the electron concentration $e/a$, and that $T_m$ increases
with increasing $e/a$~\cite{2-c,3-j}. Furthermore, alloying of
Ni-Mn-Ga with a $3d$ transition element can also considerably
increase martensitic transformation temperature~\cite{4-n,5-k}.
These observations have provoked studies of Ni-Mn-Ga aimed at the
development of a new high-temperature shape memory alloys
system~\cite{4-n,6-c,7-m,8-x}.

Recent experimental investigations~\cite{6-c,7-m,8-x} were devoted
to the study of Ni-Mn-Ga alloys with deficiency in Ga,
Ni$_{51.2}$Mn$_{31.1}$Ga$_{17.7}$ and Ni$_{54}$Mn$_{25}$Ga$_{21}$,
which undergo a martensitic transformation at $M_s = 423$~K and
533~K, respectively. A high stability of martensitic
transformation temperature during thermocycling~\cite{7-m} and a
well-defined shape memory effect~\cite{6-c,8-x} were reported for
the studied alloys. The influence of aging on martensitic
transformation of Ni-Mn-Ga high-temperature shape memory alloys
has not been studied so far. This aspect, however, is important
for potential applications of these materials.

Besides Ni-Mn-Ga alloys with deficit in Ga, a martensitic
transformation could occur at high temperatures in the alloys with
Ni excess, because substitution of Mn or Ga for Ni results in an
increase of electron concentration $e/a$. Our study of
Ni$_{50+x}$Mn$_{25-x}$Ga$_{25}$ $(4 \le x \le 9.75)$ has shown
that this is indeed the case and in the alloys with $x \ge 7.5$
the martensitic transformation is observed at temperatures above
530~K~\cite{9-k}. In the present work we report on the influence
of aging on martensitic transformation in a
Ni$_{57.5}$Mn$_{17.5}$Ga$_{25}$ composition. The obtained
experimental results indicated a drastic difference between aging
performed in the martensitic and austenitic state and showed
unusual aging-induced features of martensitic transformation,
observed so far only in Ni-Ti shape memory alloys.

\section{Experimental details}

Ingots of the above mentioned nominal composition were prepared by
a conventional arc-melting method. The ingots were annealed at
1100~K for 777.6~ks and quenched in water. Optical observation of
annealed samples showed no trace of a secondary phase. EDX
characterization confirmed that real composition of the samples
was close to the nominal one. Samples for calorimetric
measurements were cut from the middle part of the ingots.
Characteristic temperatures of direct and reverse martensitic
transformations were determined from the results of differential
scanning calorimetry (DSC) measurements, performed with a
heating/cooling rate 10~K/min.

An example of DSC scans performed on the samples before aging
procedures is shown in Fig.~1. Typical feature of all the samples
studied is the absence of a well-defined anomaly at the austenite
start temperature $A_s$. The DSC heating curve smoothly deviates
from the baseline, reaches a maximum at $A_p = 544$~K and exhibits
a marked change in the slope at austenite finish temperature $A_f
= 547$~K. Subsequent cooling from the austenitic state results in
martensitic transformation with characteristic temperatures
martensite start $M_s = 512$~K and martensite finish $M_f =
502$~K. It is worth noting that for the samples studied the
exothermic peak has a complex character, which could be attributed
to two (or more) martensitic phases forming upon cooling from the
high-temperature state.

\begin{figure}
\begin{center}
\includegraphics[width=6cm]{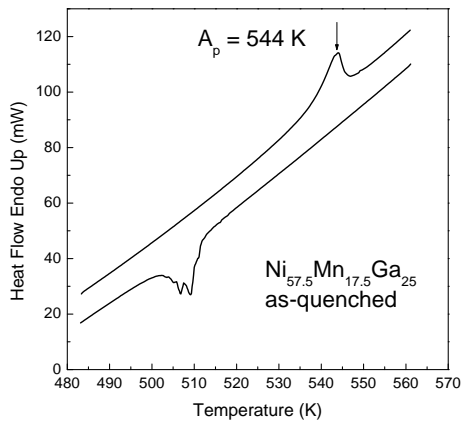}
\end{center}
\caption{Heating and cooling DSC curves of as-quenched
Ni$_{57.5}$Mn$_{17.5}$Ga$_{25}$ alloy.}
\end{figure}

\section{Results and discussion}
\subsection{Aging in the austenitic state}

Aging in the austenitic state was performed at a temperature
$T_{\mathrm{aging}} = 560$~K. Since the aging did not affect
significantly $M_s$ and $M_f$ temperatures, shown in Fig.~2 are
DSC curves of the aged samples measured at heating immediately
after each thermal treatments. These results evidence that aging
in the austenitic state results in a marked decrease of the
reverse martensitic transformation temperature (Fig.~3). As
evident from Fig.~3, the $A_p$ temperature decreases with aging
time approximately exponentially and for aging time $t \ge 432$~ks
it is equal to 520~K, a 20~K lower than that observed in the
samples before aging.

\begin{figure}
\begin{center}
\includegraphics[width=6cm]{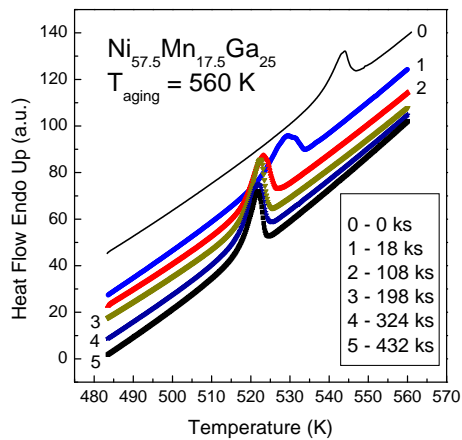}
\end{center}
\caption{Heating DSC curves of Ni$_{57.5}$Mn$_{17.5}$Ga$_{25}$
samples aged in the austenitic state at $T_{\mathrm{aging}} =
560$~K.}
\end{figure}

\begin{figure}
\begin{center}
\includegraphics[width=6cm]{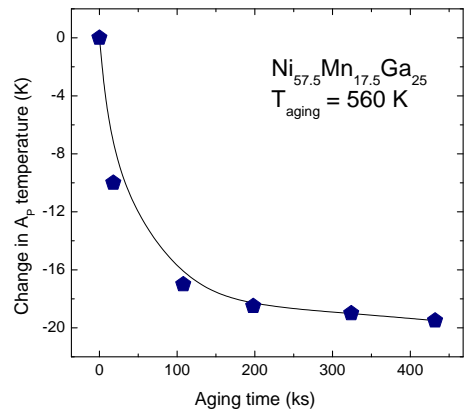}
\end{center}
\caption{Change in $A_p$ temperature after aging of
Ni$_{57.5}$Mn$_{17.5}$Ga$_{25}$ in the austenitic state as a
function of aging time.}
\end{figure}

In our opinion there are several scenarios, which could explain
the decrease of $A_p$ after aging in the austenitic state.
Considering that the samples after water quenching have a high
concentration of vacancies, which is enhanced by a significant
deviation of the studied composition from stoichiometry, the high
density of vacancies can actually produce a permanent strain in
the samples. In this sense the high density of vacancies is
equivalent to the deformation, leading to an increase in the
reverse martensitic transformation temperature~\cite{6-c,10-p}.
Aging of the samples at a high temperature above the martensitic
transformation followed by subsequent furnace cooling results in a
drastic decrease in the density of the vacancies. This process is
equivalent to the release of the applied stress, decreasing the
reverse martensitic transformation temperature. In this case one
can expect to observe a decrease in the direct martensitic
transformation temperature as well. Our experimental results have
shown, however, that the $M_s$ temperature is not affected by the
aging.

Besides this possibility, the decrease of $A_p$ can be explained
as caused by aging-induced changes in the thermodynamical
properties of the martensitic phase in such a way that the crystal
structure of martensite forming upon martensitic transformation in
the aged samples differs from that in the as-quenched samples.
Formation of the austenite-aging martensite having another crystal
structure may result in the alteration of characteristic
temperatures and the temperature hysteresis of the martensitic
transformation, i.e. the features which indeed have been observed
in the samples after aging.

\subsection{Aging in the martensitic state}

Aging of Ni$_{57.5}$Mn$_{17.5}$Ga$_{25}$ performed in the
martensitic state at $T_{\mathrm{aging}} = 528$~K revealed a
typical influence of aging of a shape memory material in the
martensitic state, i.e. an increase in the reverse martensitic
transformation temperature (stabilization of martensite) (Fig.~4).
The rate of martensite aging depends on the reduced martensitic
transformation temperature $T_R = M_s/T_m$, where $M_s$ is
martensite start temperature, $T_m$ is the melting temperature of
an alloy~\cite{11-o}. Assuming that
Ni$_{57.5}$Mn$_{17.5}$Ga$_{25}$ has a melting temperature $T_m
\sim 1400$~K~\cite{12-s}, we obtain $T_R \sim 0.37$ which should
results in a considerable aging rate. It is interesting to note
that for Ni-Mn-Ga alloys the reduced martensitic transformation
temperature is very similar to that for Au-Cd~\cite{11-o} and,
therefore, a strong aging effect could be expected even at room
temperature. In fact, this has already been observed
experimentally in stoichiometric~\cite{13-h} and
off-stoichiometric~\cite{14-h} compositions of Ni-Mn-Ga alloys.

\begin{figure}
\begin{center}
\includegraphics[width=6cm]{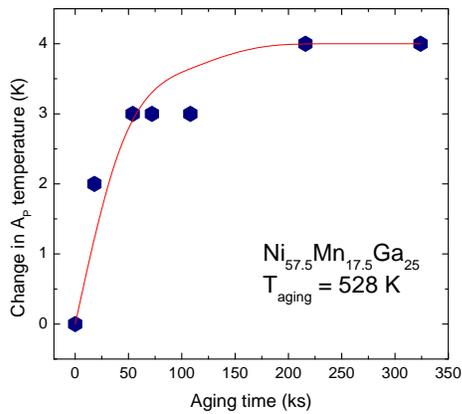}
\end{center}
\caption{Change in $A_p$ temperature after aging of
Ni$_{57.5}$Mn$_{17.5}$Ga$_{25}$ in the martensitic state as a
function of aging time.}
\end{figure}

\begin{figure}
\begin{center}
\includegraphics[width=6cm]{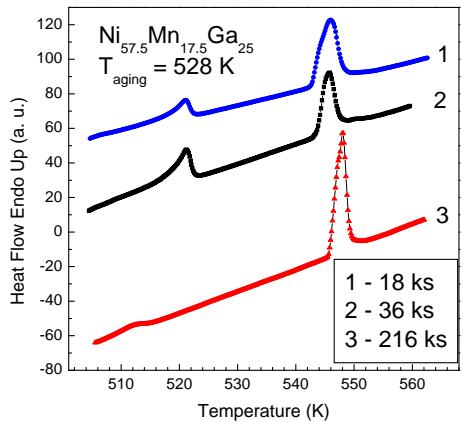}
\end{center}
\caption{Influence of aging performed in the martensitic state at
$T_{\mathrm{aging}} = 528$~K on transformation behavior of
Ni$_{57.5}$Mn$_{17.5}$Ga$_{25}$.}
\end{figure}

The most interesting and significant finding of our study is the
appearance of an additional DSC peak on heating curves of the
samples aged in the martensitic state at $T_{\mathrm{aging}} =
528$~K (Fig.~5). This well-defined two-step martensitic
transformation has been observed only for the reverse martensitic
transformation; neither martensite start temperature $M_s$ nor the
shape of the exothermic peak changed significantly after the aging
procedure. As far as we are aware, such an unusual phenomenon,
appearing after aging in the martensitic state, has been observed
so far only in Ni-Ti shape memory alloys~\cite{15-z}. This complex
transformation behavior in Ni-Ti has attracted a considerable
interest and its mechanism is still not understood clearly (see
Refs.~\onlinecite{16-b,17-k,18-d,19-k} and references therein).

Our preliminary study of aging-induced two-step martensitic
transformation in Ni$_{57.5}$Mn$_{17.5}$Ga$_{25}$ revealed that
this phenomenon depends both on aging time and aging temperature.
The additional peak, appearing at $T_I \approx 520$~K after aging
in the martensitic state at $T_{\mathrm{aging}} = 528$~K, is
already clearly seen after aging time $t = 18$~ks, although it
becomes the most pronounced in the sample aged for 36~ks. Further
increase in the aging time leads to a progressive degradation of
the peak, which transforms to a weak anomaly in the samples aged
for $t \ge 216$~ks (Fig.~5). In order to check whether the complex
transformation behavior will appear after aging at a temperature
below $T_I$, a Ni$_{57.5}$Mn$_{17.5}$Ga$_{25}$ sample was
subjected to aging at $T = 473$~K for 190.8~ks. DSC measurement of
this sample showed that after this thermal treatment the behavior
of the reverse martensitic transformation was essentially the same
as before the aging procedure.

The observed phenomenon could be explained, in principle, by
assumption that the martensitic transformation in
Ni$_{57.5}$Mn$_{17.5}$Ga$_{25}$ results in formation of a mixture
of two martensitic phases, coexisting in a wide temperature
interval. In such a case, aging at a particular temperature in the
martensitic state would promote phase separation, favoring the
martensitic phase, stable at the aging temperature. As a result of
the phase separation, two endothermic peaks, corresponding to the
martensite-martensite and martensite-austenite transformations,
might appear on the heating curve. Such a scenario, however, seems
to be inconsistent with the result of aging at $T = 473$~K, which
showed no detectable influence on the transformation behavior of
Ni$_{57.5}$Mn$_{17.5}$Ga$_{25}$.

It seems to be more likely, therefore, that the origin of the
observed phenomenon has to be looked for in
austenite-aging-induced changes in the crystal structure of
martensite in Ni$_{57.5}$Mn$_{17.5}$Ga$_{25}$. We have suggested
that such a mechanism could be responsible for the observed
decrease of $A_p$ temperature after aging in the austenitic state
(Fig.~3). In the case of aging in the martensitic state at
$T_{\mathrm{aging}} = 528$~K, the appearance of the two-step
martensitic transformation points to the formation of an
inhomogeneous martensitic state consisting of two martensitic
phases with different transformation temperatures. The fact that
the temperature of the additional endothermic peak, $T_I \approx
520$~K is essentially the same as the $A_p$ temperature in the
samples long-term aged in the austenitic state (Figs.~2,3)
supports this suggestion.

It is worth noting that, as evident from the DSC data (Fig.~1),
the existence of a trace of the austenitic phase can be reasonably
expected at the aging temperature $T_{\mathrm{aging}} = 528$~K.
Since aging at $T = 473$~K, far below austenite start temperature,
did not result in a two-step martensitic transformation, the
presence of a small amount of austenitic phase at
$T_{\mathrm{aging}} = 528$~K is probably a crucial factor for the
formation of inhomogeneous martensitic state. It can be suggested
that after aging at $T_{\mathrm{aging}} = 528$~K the part of the
austenitic phase stabilized by the aging transforms to the
austenite-aging martensite upon cooling to room temperature. Thus,
the martensitic state of the samples cooled to room temperature
after thermal treatments consists of austenite-aging martensite
phase embedded in the matrix of the as-quenched martensite. During
subsequent heating, a transformation from the austenite-aging
martensite to the stabilized austenite takes place at $T_I \approx
520$~K; at further heating, reverse transformation of the matrix
martensite occurs at $A_p \approx 545$~K.

As evident from Fig.~5, the complex transformation behavior in
Ni$_{57.5}$Mn$_{17.5}$Ga$_{25}$ depends on aging time and become
less pronounced in the samples aged for $t \ge 216$~ks. This can
be presumably ascribed to the competition between thermodynamic
stabilization of austenitic and martensitic phases coexisting at
$T_{\mathrm{aging}} = 528$~K; the latter process dominates in the
case of a long-term aging.

\section{Conclusion}

Differential scanning calorimetry study of the influence of aging
on martensitic transformation in the
Ni$_{57.5}$Mn$_{17.5}$Ga$_{25}$ shape memory alloy revealed
complex and intriguing transformation behavior. Decrease in the
reverse martensitic transformation temperature is observed after
aging in the austenitic state. The main and most interesting
finding of this work is the observation of a complex
transformation behavior after aging in the martensitic state at
$T_{\mathrm{aging}} = 528$~K. As for now, we presume that both the
decrease of $A_p$ temperature after aging in the austenitic state
and the multi-step martensitic transformation appearing after
aging in the martensitic state may be accounted for by the
aging-induced formation of a different type of martensitic phase
having a lower, as compared to the as-quenched martensite,
transformation temperature. To give an unambiguous interpretation
on the nature of this effect, further intensive studies of
Ni-Mn-Ga alloys with this or similar composition are necessary.

\section*{Acknowledgements}

This work was partially supported by the Industrial Technology
Research Grant Program in 2002 from New Energy and Industrial
Technology Development Organization (NEDO) and by the
Grant-in-Aids for Scientific Research from the Ministry of
Education, Culture, Science, Sports and Technology (MEXT), Japan.
One of the authors (VVK) gratefully acknowledges the Japan Society
for the Promotion of Science (JSPS) for a Postdoctoral Fellowship
Award.

\end{document}